\documentclass[a4paper]{cas-sc}

\usepackage[numbers]{natbib}
\usepackage{tcolorbox}
\usepackage{subfigure}
\usepackage{enumitem}
\usepackage{setspace}
\usepackage{amsmath}
\usepackage{bm}
\usepackage{ragged2e}
\usepackage{autobreak}
\usepackage{multirow}
\usepackage{listings}
\usepackage{color}
\definecolor{dkgreen}{rgb}{0,0.6,0}
\definecolor{gray}{rgb}{0.5,0.5,0.5}
\definecolor{mauve}{rgb}{0.58,0,0.82}
\def\tsc#1{\csdef{#1}{\textsc{\lowercase{#1}}\xspace}}
\tsc{WGM}
\tsc{QE}
\tsc{EP}
\tsc{PMS}
\tsc{BEC}
\tsc{DE}

\begin{document}
\let\WriteBookmarks\relax
\def\floatpagepagefraction{1}
\def\textpagefraction{.001}
\shorttitle{CLEBPI: Contrastive Learning for Bug Priority Inference}
\shortauthors{Wen-Yao Wang et~al.}

\title [mode = title]{CLeBPI: Contrastive Learning for Bug Priority Inference}                      



\author[1]{Wen-Yao Wang}[style=chinese]
\ead{wenyaowang108@gmail.com}
\address[1]{Faculty of Innovation Engineering, Macau University of Science and Technology, Macau, China}

\author[1]{Chen-Hao Wu}[style=chinese]
\ead{wuchenhao78@gmail.com}

\author[2, 3]{Jie He}
\ead{hejie1213@126.com}
\cormark[1]
\cortext[1]{Corresponding author}
\address[2]{College of Computer Science and Electronic Engineering, Hunan University, Changsha, Hunan, China}
\address[3]{Guangxi Key Laboratory of Machine Vision and Intelligent Control, Wuzhou University, Wuzhou, Guangxi, China}


\begin{abstract}
Automated bug priority inference can reduce the time overhead of bug triagers for priority assignments, improving the efficiency of software maintenance. Currently, there are two orthogonal lines for this task, i.e., traditional machine learning based (TML-based) and neural network based (NN-based) approaches. Although these approaches achieve competitive performance, our observation finds that existing approaches face the following two issues: 1) TML-based approaches require much manual feature engineering and cannot learn the semantic information of bug reports; 2) Both TML-based and NN-based approaches cannot effectively address the label imbalance problem because they are difficult to distinguish the semantic difference between bug reports with different priorities. In this paper, we propose CLeBPI (\textbf{C}ontrastive \textbf{Le}arning for \textbf{B}ug \textbf{P}riority \textbf{I}nference), which leverages pre-trained language model and contrastive learning to tackle the above-mentioned two issues. Specifically, CLeBPI is first pre-trained on a large-scale bug report corpus in a self-supervised way, thus it can automatically learn contextual representations of bug reports without manual feature engineering. Afterward, it is further pre-trained by a contrastive learning objective, which enables it to distinguish semantic differences between bug reports, learning more precise contextual representations for each bug report. When finishing pre-training, we can connect a classification layer to CLeBPI and fine-tune it for bug priority inference in a supervised way. To verify the effectiveness of CLeBPI, we choose four baseline approaches and conduct comparison experiments on a public dataset. The experimental results show that CLeBPI outperforms all baseline approaches by 23.86\%-77.80\% in terms of weighted average F1-score, showing its effectiveness.
\end{abstract}



\ExplSyntaxOn
\keys_set:nn { stm / mktitle } { nologo }
\ExplSyntaxOff

\maketitle

\section{Introduction}
\label{section1}

Bug traigers usually assign priorities for newly submitted bugs by fully understanding their corresponding bug report, which can enable developers to quickly fix the bugs with relatively high priorities, improving the efficiency of software maintenance and software quality. With the rapidly increasing number of bugs in software products, however, it takes bug traigers much time to manually assign the bug priority, affecting the bug triagers' efficiency. For example, according to \emph{Fang et al.}'s \cite{fang2021effective} statistics, there are more than 150 newly submitted bug reports in Mozilla\footnote{https://bugzilla.mozilla.org/} project everyday. In addition, except for the bug priority assignment, bug traigers also perform other software maintenance activities, such as bug severity assignment \cite{tan2020bug, tian2012information}, bug fixer assignment \cite{jonsson2016automated, zhang2017bug}, and duplicate bug detection \cite{he2020duplicate, nguyen2012duplicate}. To improve the efficiency of software maintenance, researchers have proposed some automated approaches to automatically predict the bug priority according to corresponding bug report \cite{fang2021effective, tian2015automated, umer2019cnn, umer2018emotion}. \emph{Tian et al.} \cite{tian2015automated} proposed DRONE, a linear regression-based model that uses multiple manually selected features of bug reports. Additionally, they also dealt with the label imbalance problem by a thresholding method. \emph{Umer et al.} \cite{umer2019cnn} proposed cPur, a convolutional neural network-based approach that can learn the local semantic information of bug reports and does not require manual feature engineering. \emph{Fang et al.} proposed a graph convolutional network-based approach, namely PPWGCN, which can learn the global word co-occurrence information. Moreover, they tackle the label imbalance problem by introducing the label penalty factor to the cross-entropy loss function. Hence, PPWGCN achieves state-of-the-art results in bug priority inference. Although these approaches perform well, by diving into these studies, we find that two major issues affect the effectiveness of existing approaches.

\textbf{The first issue is that traditional machine learning based (TML-based) approaches (e.g., DRONE) require much manual feature engineering and cannot learn the semantic information of bug reports.} The former means that researchers need to take much time for selecting suitable features from bug reports, thus the quality of the dataset may affect the models' effectiveness. The latter means that TML-based approaches can only learn the shallow features from bug reports, which limits their performance. For example, if two bug reports have many identical words but they have different priority labels, it is hard for TML-based approaches to effectively distinguish them. The reason is that these two bug reports are different in semantics but manually selected features are based on information retrieval techniques (e.g., BM25) and cannot capture the semantic feature.

\textbf{The second issue is that both TML-based and neural network-based (NN-based) approaches cannot effectively address the label imbalance problem because they are difficult to distinguish the semantic difference between bug reports with different priorities.} Specifically, in the Bugzilla platform\footnote{All bug reports used in our experiments are collected from this bug tracking platform.}, priority is divided into five classes, i.e., P1 to P5, where P1 denotes the highest priority while P5 denotes the lowest priority. The number of bug reports with priority P4 and P5 is much less than bug reports with other priority labels\footnote{P4 and P5 thus are regarded as the rare priority label.}, which causes the label imbalance problem in the bug report dataset. For TML-based approaches, they can hardly make accurate predictions for bug reports with priorities P4 and P5 because simple shallow features cannot effectively distinguish bug reports with rare priorities. Although DRONE introduces an extra thresholding method to address the label imbalance problem, the effect is slight. As for NN-based approaches, although they can learn the semantic information of bug reports, they can only perform effective learning for bug reports with non-rare priorities. In other words, they tend to learn the feature of the priority that have large-sized samples and ignore the feature of the priority that have small-sized samples. An obvious example is that cPur can achieve a high F1-score for predicting bug reports with the P3 label but it cannot make any correct prediction for bug reports with P4 and P5 labels.

To resolve the aforementioned two issue, we propose CLeBPI (\textbf{C}ontrastive \textbf{Le}arning for \textbf{B}ug \textbf{P}riority \textbf{P}rediction), an automated bug priority inference approach designed by the combination of the pre-trained language model \cite{devlin2018bert, lee2020biobert} and contrastive learning \cite{gao2021simcse, kim2021self}. To tackle the first issue, we utilize a Transformer-based architecture to construct CLeBBP and pre-train it on a large-scale bug report corpus by a masked language model (MLM) objective \cite{devlin2018bert} in a self-supervised way. Specifically, MLM randomly masks parts of tokens in each bug report sequence, and makes CLeBPI predict masked tokens according to their context. Hence, CLeBPI can automatically learn the contextual representation of bug reports while needing no manual feature engineering. To resolve the second issue, we perform a supervised pre-training for CLeBPI by introducing a contrastive learning objective, by which CLeBPI can effectively distinguish the semantic difference between bug reports with different priority, learning a more precise contextual representation for each bug report. Finally, we add a classification layer to CLeBPI and perform supervised fine-tuning on it for achieving automated bug priority inference.

To evaluate the effectiveness of CLeBPI, we re-use the public dataset released by \emph{Fang et al.} \cite{fang2021effective}, then choose state-of-the-art TML-based model DRONE \cite{tian2015automated} and three NN-based approaches, including cPur \cite{umer2019cnn}, word2vec \cite{mikolov2013distributed}, and state-of-the-art NN-based model PPWGCN \cite{fang2021effective}, as the baseline approaches. Specifically, we first divide this dataset into the training set, validation set, and testing set according to the previous split ratio \cite{hu2018deep}. Afterward, we use the training set to perform the pre-training for CLeBPI. When finishing pre-training, we fine-tune CLeBPI on the training set and evaluate it on the testing set. The experimental results show that CLeBPI outperforms all baseline approaches by 23.86\%-77.80\% in terms of weighted F1-score. We also analyze the effect of bug reports' length on the performance of CLeBPI, and the results show that CLeBPI can provide accurate bug priority inference for bug reports with any length.

To sum up, the major contributions of this paper are as follows:
\begin{itemize}
	\item We propose CLeBPI, a novel automated bug priority inference approach, which combines the pre-trained language model and contrastive learning to learn the precise contextual representation of bug reports with different priorities.
	\item We alleviate the label imbalance problem in bug priority inference by introducing contrastive learning pre-training for CLeBPI, which helps it effectively learn the semantic difference between bug reports.
	\item We verify the effectiveness of CLeBPI by comparing it with four baseline approaches, including state-of-the-art TML-based and NN-based approaches. Our experimental results demonstrate that CLeBPI significantly outperforms all the baseline approaches.
\end{itemize}

The remaining of this paper includes the following parts. Section \ref{section2} introduces the background knowledge of CLeBPI and our motivation. Section \ref{section3} elaborates on the framework of the proposed CLeBPI. Section \ref{section4} and \ref{section5} present the experimental setup and experimental results, respectively. Section \ref{section6} presents ablation experiments and discusses the threats to the validity of this work. Section \ref{section7} introduces the related works. Finally, Section \ref{section8} concludes the paper and points out the future research plans.

\section{Background and Motivation}\label{section2}
In this section, we introduce the background of our work, including bug reports, priority prediction, pre-trained language models, and contrastive learning. Then, we introduce the motivation for our work.

\begin{figure}
	\centering
	\includegraphics[width=1.0\linewidth]{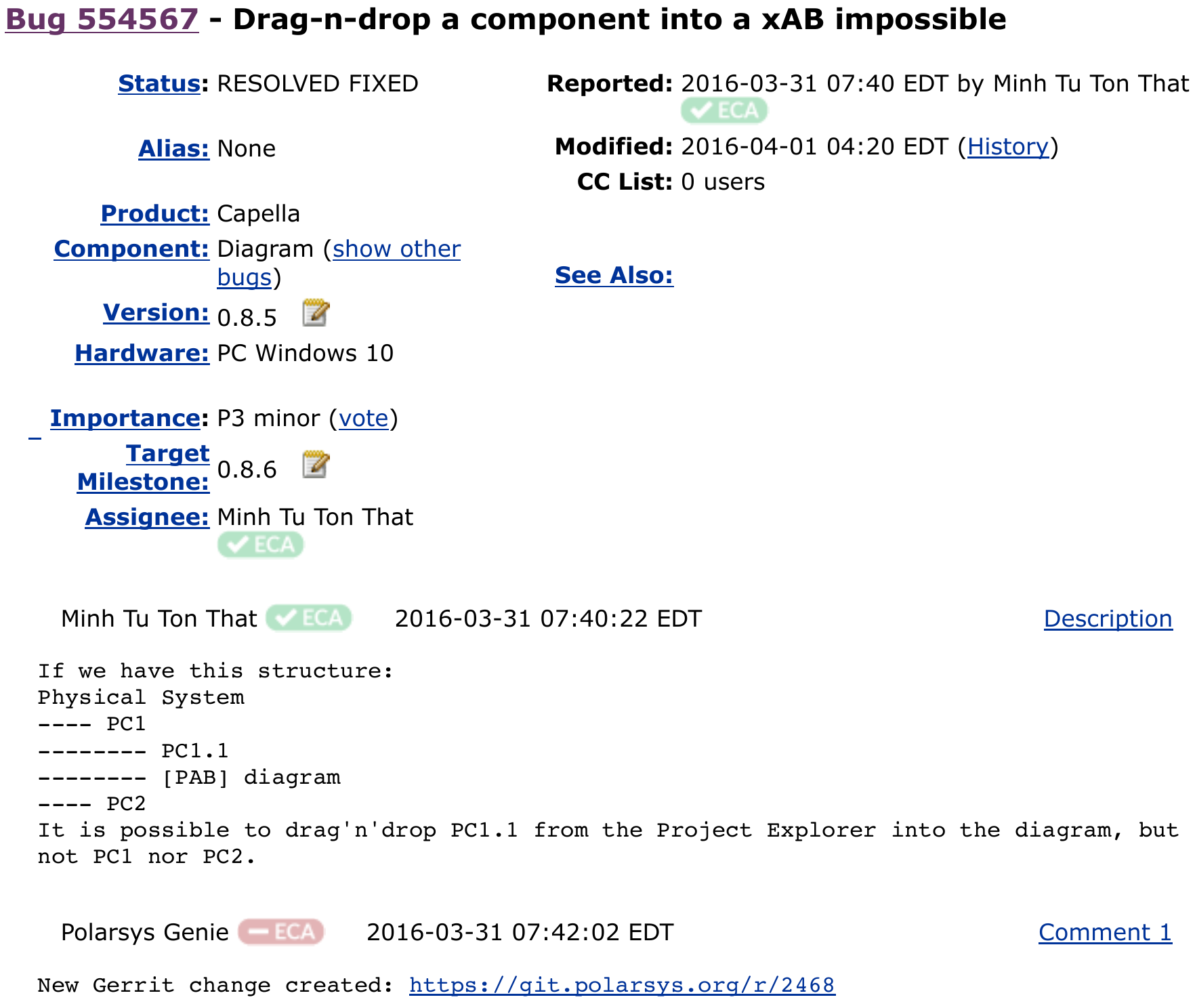}
	\caption{An example of bug report with id 554567 in Eclipse platform.}
	\label{f1}
\end{figure}

\begin{table}
	\centering
	\caption{The description of main elements in the bug report.}
	\begin{tabular}{cl}
		\toprule[1pt]
		\textbf{Element}& Description\\
		\midrule
		Summary& It laconically describes what a submitted bug is. \\
		Status& It describes the current state of a bug.\\
		Product& It indicates the project where a bug appears.\\
		Component& It is part of the Product and indicates a more specific location where a bug appears.\\
		Version& It denotes the version of the software the bug appears in.\\
		Importance& \makecell[l]{It contains two labels: one is the priority label that describes how soon the developers should \\ fix the submitted bug; another one is the severity label that describes how severe the submitted bug is.} \\
		Description& It mainly describes the reported bug specifically such as where is the bug generated.\\
		Comment& Users or developers can add their comments on the reported bug. \\
		\bottomrule[1pt]
	\end{tabular}
	\label{dbr}
\end{table}

\subsection{Bug Reports}

bug reports are submitted by developers or users to the bug tracking systems (e.g., Bugzilla, LogRocket) and contain descriptive information about the newly reported bug, such as where is the bug and what is wrong. Since different bug tracking systems may have different ways to define the bug priority \cite{saha2015these}, in this paper, we conduct research on bug reports managed by the Bugzilla platform, which is one of the most widely used bug tracking systems.

Fig.~\ref{f1} gives a bug report in the Eclipse project collected from the Bugzilla platform. From the figure, we can observe that a bug report is composed of multiple elements, such as Description, Comment, Summary, Status, Product, Component, Version, Assignee, etc. Each element has its meaning. For example, the Summary element laconically describes what the submitted bug is and the Description element specifically describes the reported bug such as where is the bug generated. A more specific introduction for the element in the bug report can be seen in Table~\ref{dbr}. Note that we only use the Description and Summary elements to perform bug priority inference since these two elements contain sufficient textual information and are more general. Although the Comment element also has the above features, it is written by other developers or users and is not always related to the submitted bug. For example, the Comment element in Fig.~\ref{f1} is simple log information. We also note that DRONE \cite{tian2015automated} utilized bug severity information to assist bug priority inference since they regard there is potential consistency between bug priority and severity. For example, a website exists some problems in some legacy browsers, such as the logo does not load and text scrambles. Since it hurts product functionality and affects user experience, its bug severity is high. However, since it only exists in legacy browsers, it has little influence on a large number of users, which means its low bug priority.

\subsection{Priority Inference}

Priority assignment is the early work in the cycle of software maintenance \cite{zhang2016towards}. When a new bug report is submitted, a bug triager first needs to ensure whether the submitted bug is a newly generated bug or enhancement \cite{fang2021effective}. Since the developer needs to deal with a large number of bug reports, the bug triager generally prioritizes the bug report before assigning it to a suitable one to resolve it. By accurately assigning bug priority, developers can fix bugs in the most efficient order, which helps improve software quality and user experience. The Bugzilla platform divides the priority into P1-P5 five different levels, where P1 is the highest fixing priority and P5 is the opposite. Although bug triagers can pre-prioritize the bug report, the growing number of bug reports makes this task time-consuming and boring. Therefore, researchers proposed some approaches to automatically recommend the bug priority for the newly submitted bug, which we described in Section~\ref{section1}.

\subsection{Pre-trained Language Model} \label{sec2.3}

Pre-trained language models \cite{devlin2018bert, peters-etal-2018-deep, yang2019xlnet} are first proposed in the natural language processing (NLP) community, which can learn the general contextual representation of words by an unsupervised pre-training on a large-scale corpus like Wikipedia. Then, the pre-trained language model can be used in different NLP tasks, such as text classification \cite{minaee2021deep}, machine translation \cite{bahdanau2016neural, sutskever2014sequence}, text summarization \cite{kouris2019abstractive}, etc, by a supervised fine-tuning \cite{devlin2018bert, yang2019xlnet}. Massive experimental results in the NLP community show that pre-trained language models have achieved state-of-the-art results in all kinds of tasks \cite{devlin2018bert}, leading the research tendency.

As the pre-trained language model becomes more popular, many domains start to privatize it \cite{beltagy2019scibert, lee2020biobert, muller2020covid}. In detail, they use the domain-specific corpus to pre-train the pre-trained language model again, making them serve a specific domain. The main reason is that the pre-trained language model pre-trained on the domain-specific corpus can learn the more precise contextual representation for domain-specific data, which further improves its performance on the domain-specific task. In the biomedical domain, for example, \emph{Lee et al.} \cite{lee2020biobert} proposed BioBERT, which is pre-trained on a large-scale biomedical corpus and achieves the state-of-the-art results in various biomedical text mining tasks such as biomedical question answering \cite{jin2022biomedical} and biomedical relation extraction \cite{wei2016assessing}. In the scientific domain, \emph{Beltagy et al.} released SCIBERT, which is pre-trained on a large multi-domain corpus of scientific publications and brings new results to a series of scientific tasks like sequence tagging \cite{huang2015bidirectional} and dependency parsing \cite{li2018seq2seq}. COVID-Twitter-BERT \cite{muller2020covid} is similar to SCIBERT, but it is pre-trained on a large corpus of Twitter messages with the topic of COVID-19.

\subsection{Contrastive Learning}\label{sec24}

The core concept of contrastive learning \cite{gao2021simcse, hadsell2006dimensionality} is to make samples that are semantically similar close together and push apart samples that are not semantically similar. According to the prior work \cite{chen2020simple}, the training objective of contrastive learning is constructed with a cross-entropy objective with in-batch negatives \cite{chen2017sampling, henderson2017efficient}. Specifically, given a set of paired examples that are semantically similar, namely $\mathcal{D}=\{(b_i, b_i^+)\}_{i=1}^n$, we assume that $r_i$ and $r_i^+$ are the representations of $b_i$ and $b_i^+$, then the training objective of contrastive learning for $(x_i, x_i^+)$ with batch size $N$ can be calculated as follows:

\begin{equation}
    \theta_i = -\log\frac{e^{sim(r_i, r_i^+)/\tau}}{\sum_{n=1}^Ne^{sim(r_i, r_n^+)/\tau}}
\end{equation}
where $sim(r_i, r_i^+)$ is the cosine similarity $\frac{r_i^\mathrm{T}r_i^+}{||r_i||\cdot||r_i^+||}$ and $\tau$ is a temperature hyperparameter. From the above equation, we can find that one critical factor of using contrastive learning is how to collect $(r_i, r_i^+)$ pairs. From the above introduction, we can note that in a mini-batch with size $N$, every sample has $N-1$ negative samples. Hence, the key challenge in contrastive learning is how to produce the positive sample for the task-specific data.

\subsection{Motivation}
Automated bug priority inference can accelerate the efficiency of software maintenance, further improving product quality and user experience. Although existing automated bug priority inference approaches (NN-based) have reported remarkable performance, we find that they only perform the accurate inference for the priority label that has large-sized samples. The major reason is that neural networks tend to learn the semantic feature of the priority label corresponding to large-sized samples. Consequently, neural networks ignore the semantic features of bug reports with the rare priority label (it only has small-sized samples). According to our investigation, all existing approaches cannot achieve more than a 10\% F1-score for the priority inference of bug reports with label P4, which also can be seen in Section~\ref{sec51}.

The above investigation result motivates us to analyze the potential reason by diving into existing approaches. By our careful analysis, we think that the existing approaches face the following two issues: 1) According to our statistics, most of bug reports (Description and Summary elements) are usually composed with source code and natural language, and a simple neural architecture cannot fully learn the contextual information of the whole bug reports; 2) the number of bug reports with rare priority label is quite few, thus existing approaches are hard to learn effective semantic features from them. These two issues make existing approaches difficult cope with the label imbalance problem.

Inspired by the success of the pre-trained language model \cite{devlin2018bert, liu2019roberta}, we think that it has the potential to help us learn the effective contextual representation of bug reports by the combination of the deep neural architecture and self-supervised pre-training. Considering that contrastive learning can boost the semantic recognition ability of pre-trained models, we can utilize contrastive learning \cite{gao2021simcse} to learn the deep semantic differences between bug reports, helping the model distinguish bug reports with different priorities. In this paper, therefore, we actively explore how to effectively combine pre-trained language models and contrastive learning, and apply them to the bug priority inference.

\section{Approach}\label{section3}
In this section, we first introduce how to build vocabulary for bug reports. Afterward, we describe the pipeline of CLeBPI, including model architecture, pre-training CLeBPI by masked language model objective, pre-training CLeBPI by contrastive learning, and training CLeBPI for bug priority inference.

\subsection{Vocabulary} \label{section3.1}

To represent bug reports by CLeBPI, we first need to map them into a set of discrete numerical sequences by a vocabulary $\mathcal{V}$. Generally, the vocabulary size is equal to the number of the unique word in the bug report corpus. However, bug report contains many compound words (e.g., FileNotFoundException and simpleconfigurator), which makes the vocabulary size become large, hurting the model learning \cite{allison2006another} since a large vocabulary exists severe data sparsity problem. Although we can control the vocabulary size by filtering low-frequency words or using top-$k$ words with the highest frequency to build vocabulary, these operations cause the out-of-vocabulary problem \cite{sennrich2015neural}, which has a negative effect on the model's performance and generalizability. According to the prior study \cite{wang2020neural}, we utilize BBPE to build the vocabulary for the bug report corpus. BBPE is based on UTF-8 encoding and has 256 basic bytes. Thus, BBPE can decompose all potential words into a set of byte n-gram, by which it can effectively control the vocabulary size.

We give an example to show how BBPE decomposes bug reports into different byte n-grams. The following is a bug report that contains Summary and Description elements.
\begin{tcolorbox}
        Pasting code with JUnit asserts prompts for static import inclusion I am using JUnit3, so my test classes extend junit.framework.TestCase. However every time when I am coping and pasting a block of text containing TestCase methods(for example assertTrue()) NetBeans asks if I want to add import clause: "import static junit.framework.Assert.assertTrue;". Not only such line is unnecessary, but it also brakes build as project source level is set to "1.4" - no static imports are available.
\end{tcolorbox}
\noindent When finishing BBPE, the bug report is transformed into a set of byte n-grams, which can be seen in the following.
\begin{tcolorbox}
        P asting Ġcode Ġwith ĠJ Unit Ġasserts Ġprompts Ġfor Ġstatic Ġimport Ġinclusion ĠI Ġam Ġusing ĠJ Unit 3 , Ġso Ġmy Ġtest Ġclasses Ġextend Ġjun it . framework . Test Case . ĠHowever Ġevery Ġtime Ġwhen ĠI Ġam Ġcoping Ġand Ġpast ing Ġa Ġblock Ġof Ġtext Ġcontaining ĠTest Case Ġmethods ( for Ġexample Ġassert True ()) ĠNet Be ans Ġasks Ġif ĠI Ġwant Ġto Ġadd Ġimport Ġclause : Ġ" import Ġstatic Ġjun it . framework . Ass ert . assert True ; ". ĠNot Ġonly Ġsuch Ġline Ġis Ġunnecessary , Ġbut Ġit Ġalso Ġbrakes Ġbuild Ġas Ġproject Ġsource Ġlevel Ġis Ġset Ġto Ġ" 1 . 4 " Ġ- Ġno Ġstatic Ġimports Ġare Ġavailable .
\end{tcolorbox}
\noindent Note that Ġ is a start marker that can be used to recover the encoded sentence. We can find that BBPE decomposes compound words into a set of byte n-grams (e.g., assertTrue->assert+True, NetBeans->Net+Be+ans).

\subsection{Model Architecture}\label{section3.2}

\begin{figure}
	\centering
	\includegraphics[width=1.0\linewidth]{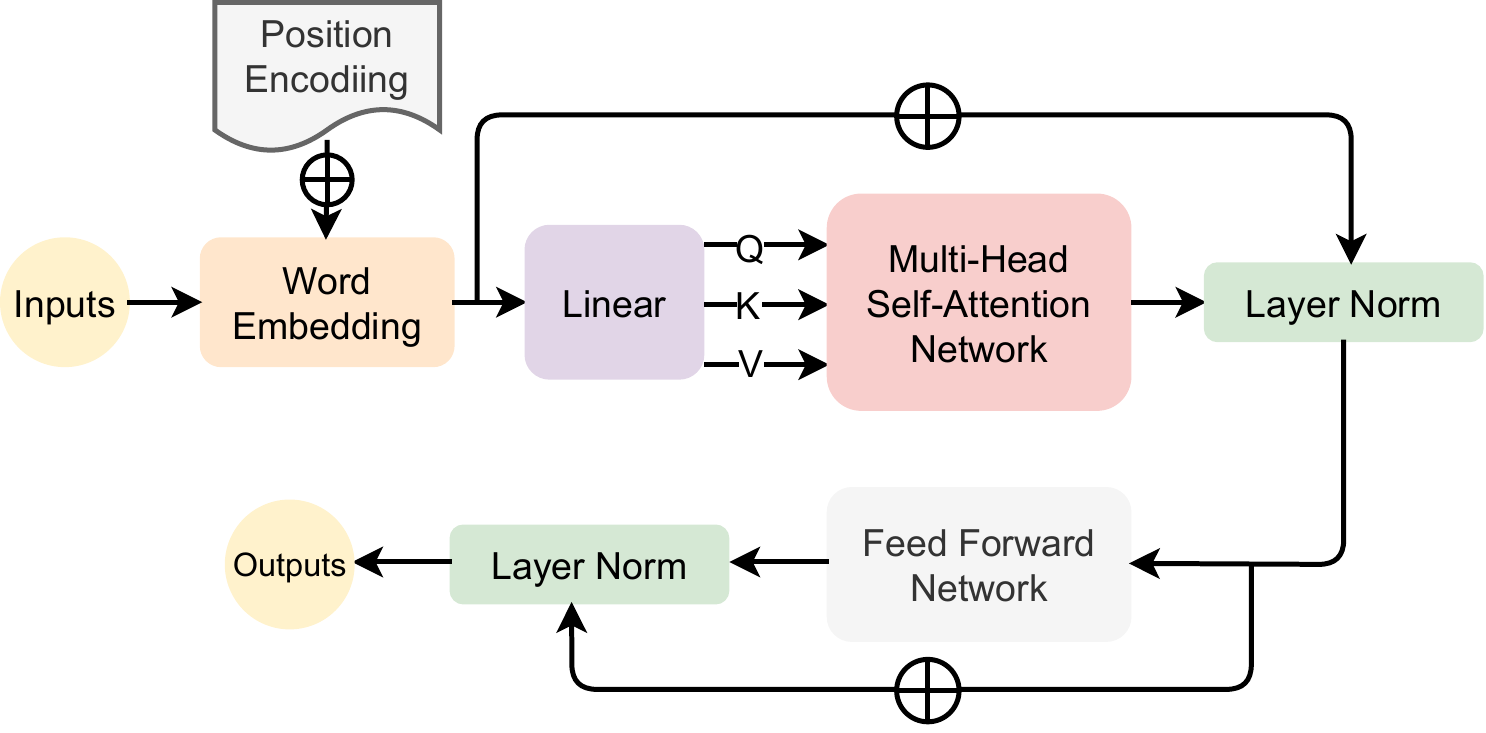}
	\caption{A single layer of Transformer encoder. Note that although we describe the word embedding and position encoding, they are not the components of the Transformer encoder layer.}
	\label{f2}
\end{figure}

We design CLeBPI by stacking Transformer \cite{vaswani2017attention} encoder layer. As shown in Fig.~\ref{f2}, it is one Transformer encoder layer, which is composed of two sub-layers, including a multi-head self-attention network and a fully connected feed-forward network. Specifically, given a bug report sequence $S = \{t_1, t_2, ..., t_l\}$ where $l$ is the sequence length, we first insert two special tokens into $S$, namely $S = \{[\text{CLS}], t_1, t_2, ..., t_l, [\text{EOS}]\}$. Before inputting the $S$ into the Transformer encoder, we need to perform a word embedding \cite{mikolov2013distributed} and a position encoding to it. For word embedding, we utilize a lookup table $\mathcal{E}\in \mathbb{R}^{d_w\times|\mathcal{V}|}$ to map each token in $S$ into a non-contextual vector $v_l\in \mathbb{R}^{d_w}$, where $d_w$ is the dimension of the word embedding and $|\mathcal{V}|$ is the vocabulary size. As for position encoding, we use a simple relative position encoding \cite{devlin2018bert} replace the absolute position encoding used in vanilla Transformer \cite{vaswani2017attention}. The major reason is that the absolute position encoding ignores the relative position information between tokens \cite{shaw2018selfattention}. More specifically, we use another lookup table $\mathcal{P}\in \mathbb{R}^{d_w\times \mathcal{L}}$, where $\mathcal{L}$ is the max length of bug reports, to map the position of token $t_l$ into a position vector $p_l\in \mathbb{R}^{d_w}$, thus CLeBPI can learn the relative position information between tokens in the training process. The input of Transformer encoder layer can be expressed as follows:
\begin{equation}
    S_i = \mathcal{E}(S) + \mathcal{P}(S)
\end{equation}

When getting the input of Transformer encoder layer, we first utilize three individual and learned linear projections to transform it into $Q$, $K$, and $V$ vectors:
\begin{equation}
    Q = S_iW_Q
\end{equation}
\begin{equation}
    K = S_iW_K
\end{equation}
\begin{equation}
    V = S_iW_V
\end{equation}
where $Q$, $K$, and $V \in \mathbb{R}^{d_m}$, and $W_Q$, $W_K$, and $W_V \in \mathbb{R}^{d_w\times d_m}$. Generally, $d_m$ is equal to $d_w$. To get the output of multi-head self-attention network, we can first calculate the output of the single-head self-attention network, which is shown as follows:
\begin{equation}
    \text{SHSAN(Q,K,V)} = \text{softmax}(\frac{QK^T}{\sqrt{d_m}})V
\end{equation}
The output of multi-head self-attention network can be expressed by the concatenation of multiple single-head self-attention network:
\begin{equation}
    \text{MHSAN}(Q,K,V) = \text{Concat}(\text{SHSAN}_1,...,\text{SHSAN}_h)W_O
\end{equation}
where $\text{Concat}$ is a concatenation operation and $\text{SHSAN}_h$ can be computed as follows:
\begin{equation}
    \text{SHSAN}_h = \text{SHSAN}(QW_h^Q,KW_h^K,VW_h^V)
\end{equation}
where the projections $W_h^Q \in \mathbb{R^{d_m\times d_q}}$, $W_h^K \in \mathbb{R^{d_m\times d_k}}$, $W_h^V \in \mathbb{R^{d_m\times d_v}}$, and $W_O \in \mathbb{R^{hd_v\times d_m}}$. Note that $d_q=d_k=d_v=d_m/h$. Next, we add a residual connection \cite{he2016deep} component to multi-head self-attention network, followed by a layer normalization \cite{ba2016layer}, thus the output can be calculated as follows:
\begin{equation}
    O_{LN} = \text{LN}(S_i+\text{MHSAN}(Q,K,V))
\end{equation}
where $O_{LN}$ is the output of the normalization layer and $\text{LN}$ denotes the layer normalization. Both residual connection and layer normalization can accelerate the convergence of the model and avoid the vanishing gradients problem. The multi-head self-attention network is connected with a fully connected feed-forward network, which is also followed by a residual connection and a layer normalization. The output of the Transformer encoder layer is computed as follows:
\begin{equation}
    \text{outputs} = \text{LN}(O_{LN}+(\text{ReLU}(O_{LN}W_1+b_1)W_2+b_2))
\end{equation}
where $W_1 \in \mathbb{R}^{d_m\times dff}$ and $W_2 \in \mathbb{R}^{dff\times d_m}$ are parameter matrices. $b_1\in \mathbb{R}^{dff}$ and $b_2\in \mathbb{R}^{d_m}$ are biases. $dff$ is equal to $4\cdot d_m$.

\subsection{Pre-training CLeBPI}\label{section3.3}

\begin{figure}
	\centering
	\includegraphics[width=1.0\linewidth]{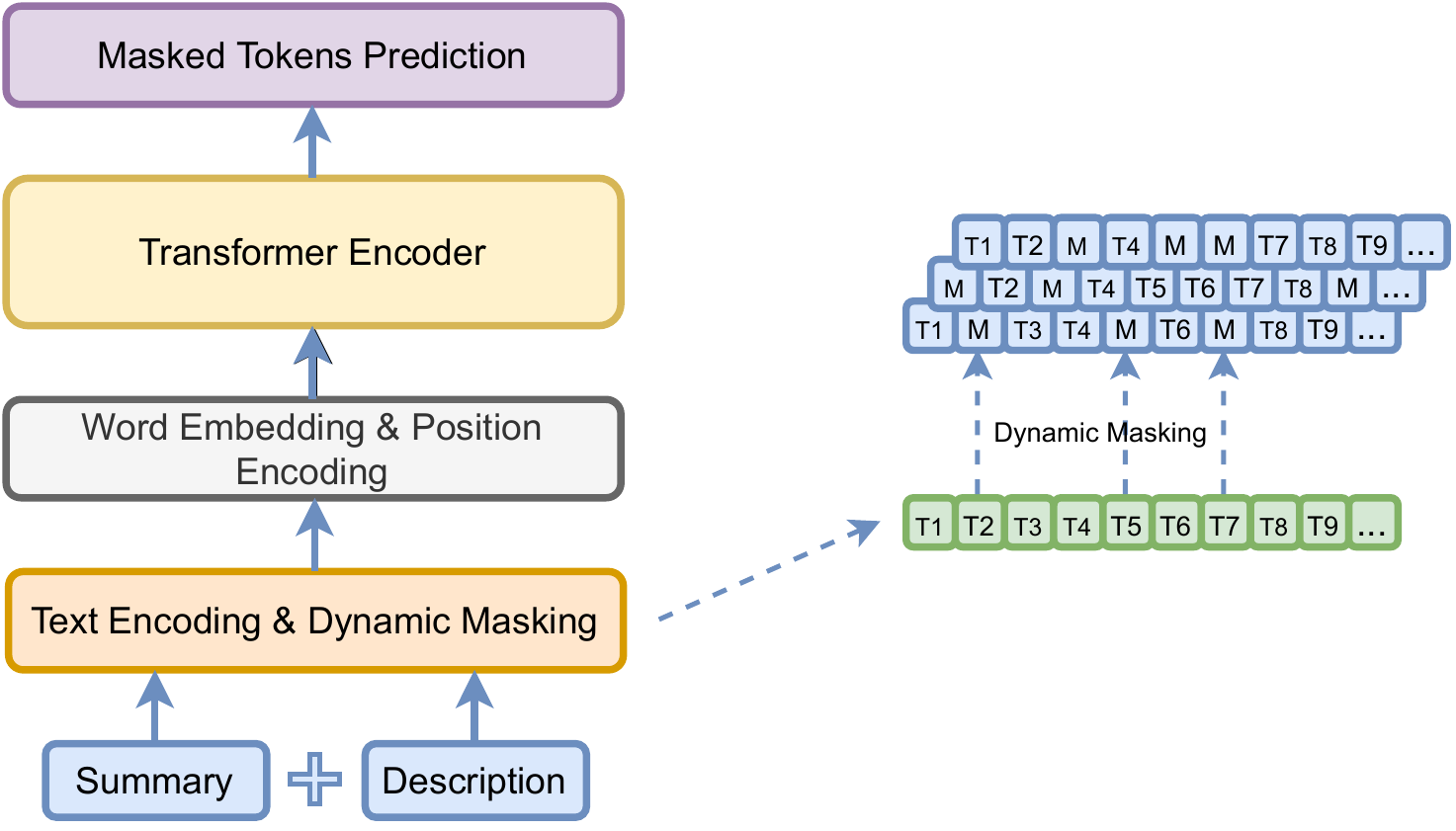}
	\caption{The left is the pipeline of pre-training CLeBPI with a masked language model, and the right is a simple example of dynamic masking.}
	\label{f3}
\end{figure}

The left of Fig.~\ref{f3} gives a pipeline of pre-training CLeBPI with a masked language model objective. We first need to perform a masking operation for each bug report sequence. Specifically, we randomly select parts of tokens in the bug report sequence and replace them with    a special $[\text{MASK}]$ token. Following the prior work \cite{devlin2018bert}, we choose 15\% of tokens in each bug report sequence and mask them in the following three ways:
\begin{itemize}
    \item Replacing them with $[\text{MASK}]$ with the 80\% probability;
    \item Replacing them with random tokens with the 10\% probability;
    \item Keeping unchanged with the 10\% probability.
\end{itemize}
Different from BERT which uses static masking, we utilize dynamic masking for each bug report sequence, which is the same as RoBERTa \cite{liu2019roberta}. As shown in the right of Fig.~\ref{f3}, dynamic masking can mask different positions for the same bug report sequence in every iteration. The advantage of this operation is that the model can thoroughly learn the contextual representation of bug reports by predicting different masked tokens. As for static masking, it only masks some stable positions of each bug report in the whole pre-training process, thus the model may tend to focus on the context of masked tokens, ignoring the context of the entire bug report. More specifically, we mask each bug report in a dynamic masking way ten times, which means that each bug report has ten variants and the scale of the original dataset becomes ten times larger.

When CLeBPI learns the representation of masked bug reports, we exploit the masked language model objective to train it by maximizing the following log-likelihood:
\begin{equation}
    \mathcal{L}_{MLM}(\theta)=\sum_{i\in \mathcal{M}}-\log p(t_i|\hat{S})
\end{equation}
where $\theta$ is the learned parameters, $\mathcal{M}$ is a set of positions of masked tokens, the probability $p(\cdot)$ is modeled by CLeBPI, $t_i$ is the masked token, and $\hat{S}$ is the remaining tokens. Compared with the standard autoregressive language model \cite{radford2018improving} that predicts the next token according to its left context, the masked language model enables the model to comprehensively consider both the left and right context of the masked token, making a reasonable inference. By the masked language model, CLeBPI can effectively learn the contextual representation of bug reports, capturing their semantic information.

\begin{table}[!t]
	\centering
	\caption{Statistics of the dataset.}
	\begin{tabular}{lll}
		\toprule[1pt]
		Dateset& Preject& Size\\
		\midrule
		L & the number of Transformer encoder layer& 12 \\
		$d_w, d_m$ & Hidden size& 768 \\
		$dff$ & Inner hidden size of feed-forward network& 3072 \\
		h & Attention heads & 12 \\
		$d_q, d_k, d_v$ & Attention head size & 64 \\
		Dropout & Dropout rate & 0.1 \\
		Attention Dropout & Dropout rate in self-attention network & 0.1 \\
		Warmup Steps & Number of steps used for a linear warmup from 0 to learning rate & 1K \\
		Learning rate & The initial learning rate for AdamW optimizer & 5e-5 \\
		Batch Size & - & 16 \\
		Weight Decay & Weight decay for AdamW optimizer & 0.01 \\
		Max Steps & Total number of training steps to perform & 275K \\
		Learning rate decay & The scheduler type to use & Linear \\
		Adam $\epsilon$ & The $\epsilon$ hyperparameter for Adam optimizer & 1e-8 \\
		AdamW $\beta_1$ & The $\beta_1$ hyperparameter for AdamW optimizer & 0.9 \\
		AdamW $\beta_2$ & The $\beta_2$ hyperparameter for AdamW optimizer & 0.999 \\
		\bottomrule[1pt]
	\end{tabular}
	\label{ds1}
\end{table}

Table~\ref{ds1} gives the details of pre-training CLeBPI with a masked language model. Following the prior work \cite{devlin2018bert, liu2019roberta}, the hyperparameter setting of CLeBPI is: $L=12$, $h=12$, $d_w=d_m=768$, $dff=3072$, and $d_q=d_k=d_v=64$. We optimize CLeBPI by a AdamW optimizer \cite{loshchilov2017decoupled} with a learning rate of 5e-5, $\beta_1=0.9$, $\beta_2=0.999$, L2 weight decay of 0.01, and a linear decay of the learning rate. We set the batch size and max length of bug report sequence to 16 and 512, respectively. We pre-train CLeBPI 20 epochs, which is equal to about 275,000 training steps. We initialize CLeBPI with the weight of RoBERTa \cite{liu2019roberta}, which is the same as CodeBERT \cite{feng2020codebert} and BioBERT \cite{lee2020biobert}.

\subsection{Pre-training CLeBPI with Contrastive Learning}\label{sec3.4}

\begin{figure}[t]
	\centering
	\includegraphics[width=1.0\linewidth]{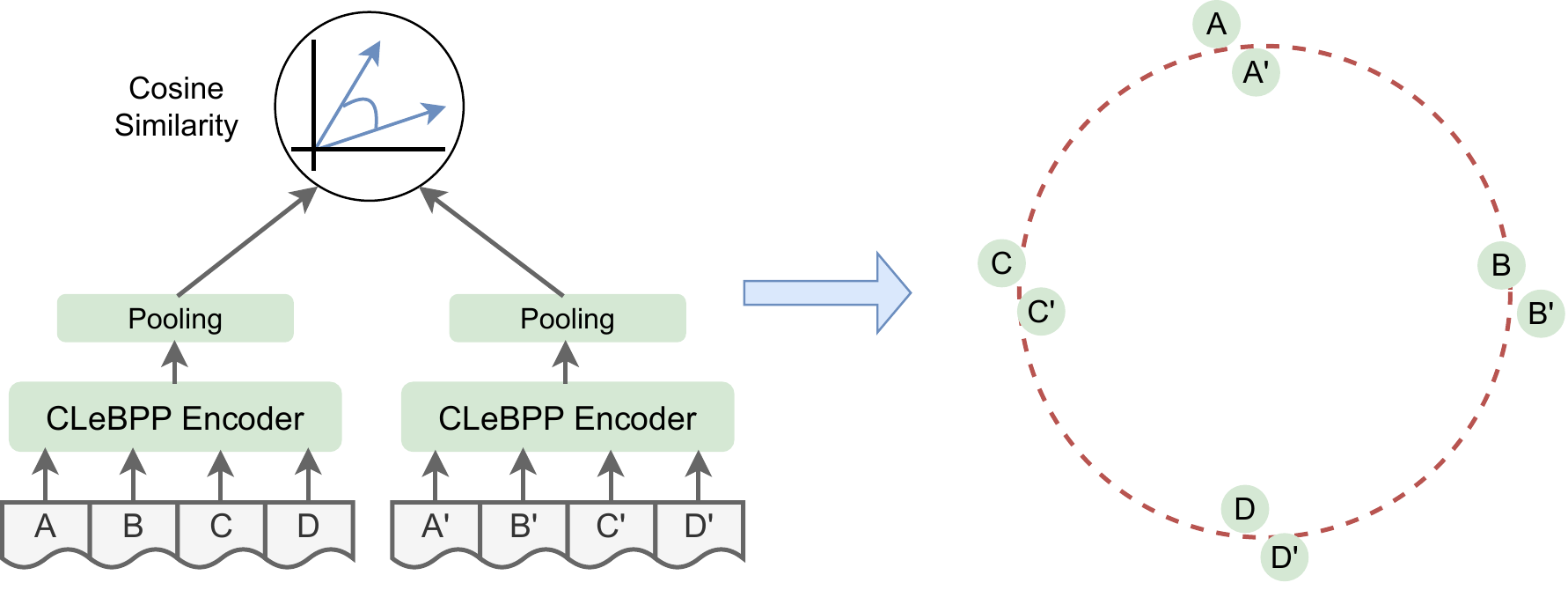}
	\caption{The left is a pipeline of pre-training CLeBPI with contrastive learning, and the right is an example to show the effect of contrastive learning.}
	\label{f4}
\end{figure}
The left of Fig.~\ref{f4} gives a pipeline of pre-training CLeBPI with contrastive learning objective \cite{gao2021simcse, huang2021cosqa}. As we describe in Section~\ref{sec24}, the major challenge of contrastive learning is how to produce positive instances for the given bug report (denote it as $\text{A}$). Inspired by the masked language model, we can produce positive instances for the given bug report by randomly replacing one token in it with a $[\text{MASK}]$ token (denote the generated positive instance as $\text{A}'$). According to the previous studies \cite{gao2021simcse, huang2021cosqa}, a minor modification to a textual sequence has little influence on its semantics. In this situation, $\text{A}$ and $\text{A}'$ are semantically similar since they just have one different token. Hence, we can regard this augmented bug report $\text{A}'$ as a positive instance of bug report $\text{A}$. Additionally, considering that some slight grammar errors cannot affect reading and understanding, we also use two extra methods to generate positive instances for the given bug report: 1) randomly delete one word in the original bug report; 2) randomly exchange the positions of two words in the original bug report. In our experiments, we produce the positive instance for the given bug report by exchanging the positions of two words in it because this method can bring the highest performance improvement. We also make a comprehensive comparison of these three methods in Section~\ref{section6}.

Formally, we can use the above method to build a set of semantically similar bug report pairs, denoting them as $\mathcal{B}=\{(s_i, s_i^+)\}$. Afterward, we can generate negative samples by in-batch augmentation method \cite{gao2021simcse, huang2021cosqa, husain2019codesearchnet}. Suppose that the batch size is $N$, which means that each batch contains $N$ semantically similar bug report pairs. For each pair $(s_i, s_i^+)$, we can pair $s_i$ with $N-1$ positive instances generated from other bug reports, and regard the $N-1$ pairs as the negative instances for pair $(s_i, s_j^+)$. The training objective of contrastive learning is to minimize $\ell_i$:
\begin{equation}
    \ell_i = -\log\frac{e^{\text{sim}(\textbf{r}_i,\textbf{r}_i^+)/\tau}}{\sum_{j=1}^N e^{\text{sim}(\textbf{r}_i,\textbf{r}_j^+)/\tau}}
\end{equation}
where $\text{sim}(\cdot)$ is the cosine similarity, $r_i$, $r_i^+$, and $r_j^+$ are the contextual representations of $s_i$, $s_i^+$, and $s_j^+$, and $\tau$ is temperature factor. As shown in the right of Fig.~\ref{f4}, we give an example to elaborate on the function of contrastive learning. It makes the semantically similar samples close together (e.g., A and A$'$, B and B$'$, etc) and the semantically dissimilar samples apart from each other (e.g., C and D, C$'$ and D). In other words, contrastive learning enables CLeBPI to learn the semantic difference between bug reports, which helps it distinguish bug reports with different priorities. From a representation learning perspective, contrastive learning works because it can alleviate the \textit{anisotropy} problem in the representation modelled by pre-trained language models \cite{ethayarajh2019contextual, li2020sentence}. That is, the learned embedding by masked language model objective occupies a narrow cone in the vector space, in which two semantically dissimilar vectors are close with each other. As a result, the model cannot distinguish the semantic difference between vectors. Due to the property of contrastive learning (i.e., makes the negative instances apart), it can alleviate the \textit{anisotropy} problem by improving the uniformity of vector space \cite{gao2021simcse}.

\paragraph{Pre-training details} The most hyperparameter settings in pre-training CLeBPI by contrastive learning objective are the same as that in pre-training CLeBPI by masked language model objective. We mainly change the batch size, learning rate, and the number of the training epoch to 32, 3e-5, and 5, respectively. Before pre-training CLeBPI by contrastive learning objective, we initialize it with the weight of CLeBPI pre-trained by the masked language model objective.

\subsection{Training CLeBPI for Bug Priority Inference}

\begin{figure}
	\centering
	\includegraphics[width=0.5\linewidth]{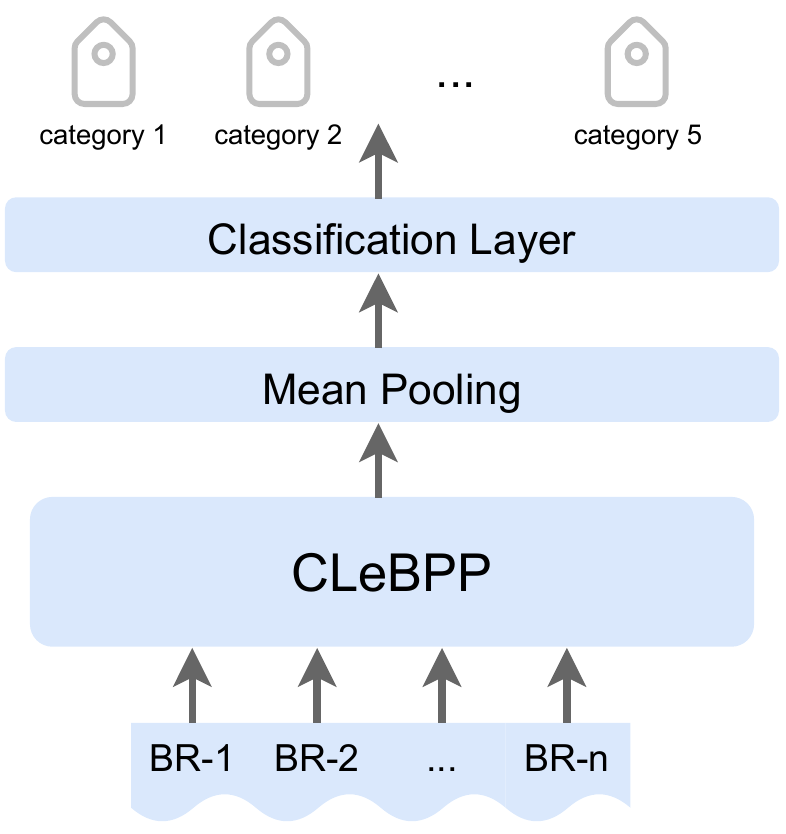}
	\caption{CLeBPI for bug priority inference.}
	\label{f5}
\end{figure}
When finishing the two-stage pre-training of CLeBPI, we can utilize pre-trained CLeBPI to perform bug priority inference. Fig.~\ref{f5} shows how to train CLeBPI for bug priority inference. Given a batch of bug reports $\mathcal{X}=\{\text{bug report}_1, ..., \text{bug report}_n\}$, where $n$ is the batch size, we input it to the CLeBPI and obtain the corresponding output:
\begin{equation}
    O_x = \text{CLeBPI}(\mathcal{X})
\end{equation}
Then, we input $O_x$ to an aggregation layer and perform a mean pooling to it, getting the contextual representation of each bug report.
\begin{equation}
    O_{mp} = \text{meanpooling}(O_x)
\end{equation}
Finally, we feed $O_{mp}$ into a classification layer to predict the priority label $c$ of each bug report:
\begin{equation}
    p(c|O_{mp})=\text{softmax}(WO_{mp})
\end{equation}
where $W$ is a parameter matrix and $\text{softmax}$ is a classifier. We train the parameters of CLeBPI and $W$ by maximizing the log-probability of the correct label.

\paragraph{Training details} In the training phase, most hyperparameter settings are the same as that in the pre-training phase. We mainly change the batch size, learning rate, and the number of the training epoch to 64, 5e-6, and 10, respectively. We also limit the max length of bug reports to 256 because CLeBPI can get the best performance with this setting. We also discuss the effect of the max length of bug reports in Section~\ref{sec53}.

\section{Experiment Setup}\label{section4}

\subsection{Research Questions} \label{sec41}
Our work mainly explore the following three research questions (RQ):

\begin{tcolorbox}
	\textbf{RQ1:} How effective is CLeBPI when compared with (1) the state-of-the-art TML-based approach, and (2) the state-of-the-art NN-based approaches?
\end{tcolorbox}
In RQ1, we mainly explore the effectiveness of CLeBPI, which includes two parts: one is whether CLeBPI performs better than the existing approaches, and another one is whether CLeBPI can effectively address the label imbalance problem by comparison with the existing approaches. The former can demonstrate that CLeBPI can learn the effective contextual representation of bug reports compared with the existing approaches, and the latter can verify that contrastive learning can help CLeBPI alleviate the label imbalance problem by learning the deep semantic differences between bug reports. Thus, we choose one state-of-the-art TML-based approach and three effective NN-based approaches (including the state-of-the-art approach), then compare CLeBPI with them on an open-source dataset, to evaluate its performance.

\begin{tcolorbox}
	\textbf{RQ2}: How effective is CLeBPI when compared with the pre-trained language model in other fields (e.g., BERT in the NLP field)?
\end{tcolorbox}
In RQ2, we mainly investigate whether CLeBPI is more effective than some pre-trained language models in other fields. The reason we perform such research is that CLeBPI is also pre-trained by masked language model objective and contrastive learning objective, we thus think that it is necessary to compare CLeBPI with existing pre-trained language models. Additionally, existing pre-trained language models claim that they can achieve competitive results on various of downstream tasks by a supervised fine-tuning. If CLeBPI performs better than these pre-trained language models, it can further support the effectiveness of CLeBPI.

\begin{tcolorbox}
	\textbf{RQ3}: How does the length of bug report affect the performance of CLeBPI?
\end{tcolorbox}
In RQ3, we mainly explore (1) whether bug reports with different lengths affect the performance of CLeBPI, and (2) whether the choice of the max length of bug report affects the performance of CLeBPI. The reason to perform the former exploration is that some previous studies \cite{tan2020bug, zhang2017bug, zhang2016towards} find that short bug reports may have a negative impact on models' performance since they only contain limited textual information. Considering that CLeBPI is proficient in learning contextual representation and semantic differences of bug reports, we tend to explore whether CLeBPI can eliminate this negative impact. As for the latter research, we mainly explore how the choice of bug reports' max length affects the performance of CLeBPI, including the effectiveness and training time. To perform the first research, we divide bug reports into 6 intervals by the length with strides of 100, including 0-100, 100-200, 200-300, 300-400, 400-500, and more than 500. Then, we calculate the inference accuracy of CLeBPI in each length interval and observe whether CLeBPI has different performance in each length interval. To perform the second research, we choose different bug reports' max length settings when we train CLeBPI for the bug priority inference, i.e., 64, 128, 256, and 512. Then, we calculate the inference accuracy and training time of CLeBPI under each max length setting.

\subsection{Dataset and Baselines}\label{section4.2}

\begin{table}
	\centering
	\caption{The statistics of the bug report in each project of the Bugzilla platform.}
	\begin{tabular}{ccc}
		\toprule[1pt]
		Project& Number of Bug Reports& Average Length\\
		\midrule
		Mozillag& 112,750& 142.61\\
		Eclipse& 106,627& 114.13\\
		Netbeans& 23,236& 200.15\\
		GCC& 33,026& 229.21\\
		\midrule
		Overall& 275,639 & 171.53\\
		\bottomrule[1pt]
	\end{tabular}
	\label{dss1}
\end{table}

\begin{table}[t]
	\centering
	\caption{The statistics of the bug report dataset that is used in the bug priority inference.}
	\begin{tabular}{l|c|c|c}
		\toprule[1pt]
		\multirow{2}*{\textbf{-}} & \multicolumn{3}{c}{Number of Bug Reports} \\
		\cline{2-4}
		& Training set & Valid set & Test set \\
		\midrule
		Bug Priority Inference & 176,588 & 21,939 & 21,909 \\
		\bottomrule[1pt]
	\end{tabular}
	\label{dss2}
\end{table}

\begin{table}[t]
	\centering
	\caption{The statistics of bug reports according to the priority label.}
	\begin{tabular}{l|c|c|c}
		\toprule[1pt]
		\multirow{2}*{\textbf{-}} & \multicolumn{3}{c}{Number of Priority Labels} \\
		\cline{2-4}
		& Training set & Valid set & Test set \\
		\midrule
		P1 & 34,544 & 4,375 & 4,344 \\
		P2 & 32,569 & 4,031 & 3,960 \\
		P3 & 102,634 & 12,703 & 12,782 \\
		P4 & \textbf{2,935} & \textbf{352} & \textbf{350} \\
		P5 & \textbf{3,906} & \textbf{478} & \textbf{473} \\
		\bottomrule[1pt]
	\end{tabular}
	\label{imbalance}
\end{table}

\subsubsection{Dataset} \label{sec421}
We conduct all the experiments in this work on the open-source dataset released by \emph{Fang et al.} \cite{fang2021effective}. Specifically, they collected more than 270,000 bug reports from four open-source projects in the Bugzilla platform, i.e., Mozilla, Eclipse, Netbeans, and GNU compiler collection (GCC), from Feb. 2000 to Sep. 2020. As shown in Table~\ref{dss1}, it gives the detailed statistics of bug reports in each project. For each bug report, we only reserve its Summary and Description elements, then combine these two elements and use the Summary element as the first sentence. Moreover, we have no pre-processes for bug reports because BBPE can effectively encode each token in bug reports. Although bug reports contain other elements, we think these two elements can provide enough sufficient textual information to be learned by CLeBPI. Afterward, following the prior work \cite{fang2022prhan, hu2018deep}, we divide 80\% of all bug reports into the training set, 10\% of all bug reports into the validation set, and the remaining bug reports into the testing set, respectively. We use the training set to perform the first-stage pre-training for CLeBPI and evaluate it on validation and testing sets. To perform the second-stage pre-training, we build a new dataset based on the dataset used for the first stage. We conduct a data augmentation for each bug report by randomly exchanging the positions of two tokens in it, to generate a positive instance for each bug report, then we combine them into a tuple as the input in the second-stage pre-training.

When finishing pre-training, we need to further process the dataset used in the first-stage pre-training for the following training of bug priority inference. Specifically, we filter bug reports that have no priority label because the training process for bug priority inference is supervised and requires labeled data. The statistics of labeled bug reports are shown in Table~\ref{dss2}. In addition, there exists a severe label imbalance problem in the bug report dataset. As shown in Tabel~\ref{imbalance}, from the table we can observe that bug report with priority P3 occupies most parts of the whole dataset and are about 30 times more than bug report with priority P4 or P5, causing the label imbalance problem. The label imbalance problem may make the model focus on learning features of the bug report whose priority is P3 and ignore the features of bug reports with priority P4 or P5. Consequently, the model has extremely high accuracy on the priority inference of bug reports with label P3 while cannot make any effective inference for the bug report whose priority is P4 or P5.

\subsubsection{Baselines}

\paragraph{Baselines in RQ1} We choose four baseline approaches and compare them with CLeBPI. In detail, we choose one state-of-the-art TML-based approach, namely DRONE \cite{tian2015automated}, which is a linear regression model that uses multiple factors of bug reports and deals with the label imbalance problem by a thresholding method. Additionally, we choose three NN-based approaches, i.e., word2vec \cite{mikolov2013distributed}, cPur \cite{umer2019cnn}, and PPWGCN \cite{fang2021effective}, all of which are word embedding, convolutional neural network, and graph convolutional network based approaches. Especially, PPWGCN is the state-of-the-art bug priority inference approach.

\paragraph{Baselines in RQ2} We choose two famous pre-trained language models in the NLP field as the baseline approaches, namely BERT \cite{devlin2018bert} and RoBERTa \cite{liu2019roberta}, all of which have achieved the state-of-the-art results on all kinds of NLP tasks. Besides, we also choose CodeBERT \cite{feng2020codebert}, a pre-trained language model used for the representation of source code and natural language, as another baseline approach. We compare CLeBPI against them to verify its effectiveness. 

\subsection{Evaluation Metrics}
Following the prior bug priority inference approaches \cite{fang2021effective, tian2015automated, umer2019cnn}, we choose accuracy (A), precision (P), recall (R), and F1-score (F) as the evaluation metrics, and we can define these metrics by the following equations:
\begin{equation}
    A = \frac{\text{Number of labels that are predicted correctly}}{\text{The total number of labels}}
\end{equation}
\begin{equation}
    P_i = \frac{TP}{TP+FP}, R_i = \frac{TP}{TP+FN}, F_i = \frac{2\times P_i\times R_i}{P_i+R_i}
\end{equation}
where $TP$ (true positive) is the number of bug reports with priority $i$ that are predicted correctly, $FP$ is the number of bug reports with other priority but are predicted as priority $i$, and $FN$ is the number of bug reports with priority $i$ but are predicted as other priority. In order to alleviate the influence of label imbalance on the model performance, we also calculate the weighted average for all priority classes:
\begin{equation}
    P_{wa} = \sum_{i\in \mathcal{P}} \frac{N_i}{N_t}\times P_i, R_{wa} = \sum_{i\in \mathcal{P}} \frac{N_i}{N_t}\times R_i, F_{wa} = \sum_{i\in \mathcal{P}} \frac{N_i}{N_t}\times F_i
\end{equation}
where $\mathcal{P}$ is a set of priority classes, $N_i$ is the number of bug reports with priority $i$, and $N_t$ is the total number of bug reports. Especially, the weighted average F1-score is not always between $P_{wa}$ and $r_{wa}$\footnote{https://scikit-learn.org/stable/modules/generated/sklearn.metrics.precision\_recall\_fscore\_support.html}.

\subsection{Experiment Settings}
We conduct all experiments on a deep learning server that contains two Intel Xeon 2.20GHz CPUs, 256GB memory, and two NVIDIA Tesla V100 GPUs with 16GB memory. To implement CLeBPI, we use the following python packages: PyTorch V.1.6.0, transformers V.4.17.0 \cite{wolf2020transformers}, NumPy V.1.16.3 \cite{harris2020array}, and datasets V.1.17.0 with GPU support. The CUDA version we use is 10.2. For the baseline approaches in RQ1, if the authors have released the source code, we directly re-run the source code on our built dataset. For the baseline approaches that do not release the source code, we re-implement them according to the corresponding literature. Particularly, we keep all parameter settings constant when re-running the baseline approaches. For the baseline approaches in RQ2, we download all pre-trained language models from transformers hub\footnote{https://huggingface.co/models} and fine-tune them on our built dataset. We use API \texttt{sklearn.metrics.precision\_recall\_fscore\_support} in scikit-learn package to calculate all the evaluation metrics.

\section{Experimental Results}\label{section5}
In this section, we present the experimental results and answer the three research questions proposed in Section~\ref{sec41}.

\begin{table}[t]
	\renewcommand\arraystretch{1.4}
	\setlength\tabcolsep{3pt}
	\centering
	\caption{The performance comparison of baseline approaches and CLeBPI on the bug priority inference.}
	\begin{tabular}{c|l|cccccc}
		\toprule[1pt]
		\multirowcell{2}{Model} & \multirowcell{2}{Metric} & \multicolumn{6}{c}{Bug Priority Inference} \\
		\cline{3-8}
		&&P1 &P2 &P3 &P4 &P5 &Weighted Average \\
		\midrule[1pt]
		\multirowcell{4}{CLeBPI} & Precision (\%) & 66.87 & 56.91 & 84.12 & 36.78 & 77.65 & \textbf{74.89} \\
		& Recall (\%) & 69.81 & 50.77 & 88.11 & 23.71  & 64.12 & \textbf{76.19} \\
		& F1-score (\%) & 68.31 & 53.66 & 86.07 & 28.83 & 70.24 & \textbf{75.53} \\
		\cline{3-8}
		& Accuracy (\%) & \multicolumn{6}{c}{\textbf{76.19}} \\
		\midrule
		\multirowcell{4}{DRONE} & Precision (\%) & 50.52 & 29.45 & 48.48 & 1.43 & 25.01 & 44.19 \\
		& Recall (\%) & 55.05 & 24.98 & 43.37 & 8.73  & 1.16 & 40.90 \\
		& F1-score (\%) & 52.69 & 27.03 & 45.78 & 2.46 & 2.22 & 42.48 \\
		\cline{3-8}
		& Accuracy (\%) & \multicolumn{6}{c}{40.90} \\
		\midrule
		\multirowcell{4}{word2vec} & Precision (\%) & 49.56 & 23.22 & 60.08 & 1.28 & 35.87 & 49.87 \\
		& Recall (\%) & 18.01 & 12.23 & 88.79 & 0.22  & 43.79 & 58.53 \\
		& F1-score (\%) & 26.42 & 16.02 & 71.67 & 0.38 & 39.44 & 53.85 \\
		\cline{3-8}
		& Accuracy (\%) & \multicolumn{6}{c}{58.53} \\
		\midrule
		\multirowcell{4}{cPur} & Precision (\%) & 60.34 & 63.45 & 65.98 & 0.00 & 0.00 & 61.93 \\
		& Recall (\%) & 29.32 & 8.21 & 90.44 & 0.00  & 0.00 & 60.06 \\
		& F1-score (\%) & 39.46 & 14.54 & 76.29 & 0.00 & 0.00 & 60.98 \\
		\cline{3-8}
		& Accuracy (\%) & \multicolumn{6}{c}{60.06} \\
		\midrule
		\multirowcell{4}{PPWGCN} & Precision (\%) & 57.43 & 37.52 & 80.16 & 3.39 & 48.21 & 66.03 \\
		& Recall (\%) & 58.88 & 40.73 & 59.42 & 41.28  & 55.98 & 55.57 \\
		& F1-score (\%) & 58.15 & 39.06 & 68.25 & 6.27 & 51.81 & 60.35 \\
		\cline{3-8}
		& Accuracy (\%) & \multicolumn{6}{c}{55.57} \\
		\bottomrule[1pt]
	\end{tabular}
	\label{rq1result}
\end{table}

\subsection{Answer to RQ1: Retrieval Performance} \label{sec51}
\noindent{\textbf{The comparison of effectiveness}} Table~\ref{rq1result} gives the effectiveness comparison between CLeBPI and the baseline approaches. We calculate the accuracy (A), precision (P), recall (R), F1-score (F), and weighted average P, R, and F to measure each approach. Specifically, in terms of weighted average F1-score, CLeBPI outperforms DRONE, word2vec, cPur, and PPWGCN by 77.80\%, 40.26\%, 23.86\%, and 25.15\%. In terms of accuracy, CLeBPI outperforms DRONE, word2vec, cPur, and PPWGCN by 86.28\%, 30.17\%, 26.86\%, and 37.11\%. All these results support the effectiveness of CLeBPI. More specifically, DRONE is the worst performing approach, and the potential reason is that traditional machine learning algorithms cannot capture the semantic information of bug reports, which hinders them to perform effective priority inference for each bug report. Besides, the generalization of manual features is limited. The other three baseline approaches are based on neural networks and can automatically capture the semantic information of bug reports, thus performing better than DRONE. However, word2vec is the worst performing NN-based approach because the word embedding technique cannot learn contextual information of the whole bug report. Although cPur has a higher weighted average F1-score than PPWGCN, we can observe that cPur almost tends to learn the features of bug reports with priority P3 and cannot make any effective inference for bug reports with priority P4 or P5. By contrast, PPWGCN has a more balanced performance than cPur and makes effective inferences for bug reports with priority P5. Since we perform two-stage pre-training for CLeBPI with masked language model objective and contrastive learning objective respectively, CLeBPI can learn the precise contextual representation of bug reports and capture their semantic differences. As a result, CLeBPI performs better than all baseline approaches at each priority label inference.

\noindent{\textbf{Retrieval label imbalance}} As we described in Section~\ref{sec421}, there exists a severe label imbalance problem in the dataset for bug priority inference (i.e., P4 and P5 are rare labels), which is also shown in Table~\ref{imbalance}. From Table~\ref{rq1result} we can find that all baseline approaches cannot effectively address the label imbalance problem. Specifically, DRONE deals with the label imbalance by introducing a thresholding method but the effect is slight because DRONE is unable to perform accurate inference for bug reports whose priority is P4 or P5. As for word2vec and cPur, they perform well on priority inference for bug reports with priority P3 because it occupies the most parts of the whole bug reports and neural networks are easy to learn the semantic features of bug reports with priority P3. However, these two approaches thus are difficult to learn the semantic features of bug reports with other priorities and perform badly on priority inference for them. Different from the above two NN-based approaches, PPWGCN tackles the label imbalance by introducing a weighted loss function. Although it has the close weighted F1-score with cPur, PPWGCN performs well on the priority inference for bug reports whose priorities belong to P1, P3, and P5. In other words, PPWGCN can alleviate the label imbalance to a certain degree. Since we use a contrastive learning objective to perform an extra pre-training for CLeBPI, it can effectively distinguish the semantic differences between bug reports. Thence, CLeBPI effectively copes with the label imbalance problem. In detail, for the priority inference of bug reports with the P4 label, CLeBPI gets the absolute improvement by 22.56\% in terms of F1-score by comparison with the state-of-the-art baseline approach PPWGCN. As for the priority inference of bug reports with the P5 label, CLeBPI also outperforms the state-of-the-art result by 18.43\% in terms of F1-score. These results fully verify that CLeBPI is able to effectively alleviate the label imbalance problem.

\begin{tcolorbox}
	\textbf{Answer to RQ1}: By comparing the performance of CLeBPI with the baseline approaches, we find that CLeBPI achieves the state-of-the-art results on bug priority inference, which supports the effectiveness of CLeBPI. Additionally, we also find that CLeBPI can effectively alleviate the label imbalance problem compared with all the baseline approaches, verifying the effectiveness of contrastive learning.
\end{tcolorbox}

\begin{table*}[t]
	\renewcommand\arraystretch{1.4}
	\setlength\tabcolsep{3pt}
	\centering
	\caption{The performance comparison of CLeBPI and pre-trained language models on bug priority inference.}
	\begin{tabular}{c|l|cccccc}
		\toprule[1pt]
		\multirowcell{2}{Model} & \multirowcell{2}{Metric} & \multicolumn{6}{c}{Bug Priority Inference} \\
		\cline{3-8}
		&&P1 &P2 &P3 &P4 &P5 &Weighted Average \\
		\midrule[1pt]
		\multirowcell{4}{CLeBPI} & Precision (\%) & 66.87 & 56.91 & 84.12 & 36.78 & 77.65 & \textbf{74.89} \\
		& Recall (\%) & 69.81 & 50.77 & 88.11 & 23.71  & 64.12 & \textbf{76.19} \\
		& F1-score (\%) & 68.31 & 53.66 & 86.07 & 28.83 & 70.24 & \textbf{75.53} \\
		\cline{3-8}
		& Accuracy (\%) & \multicolumn{6}{c}{\textbf{76.19}} \\
		\midrule
		\multirowcell{4}{BERT} & Precision (\%) & 62.43 & 47.01 & 79.32 & 18.99 & 63.21 & 68.82 \\
		& Recall (\%) & 60.65 & 44.09 & 82.79 & 12.21  & 52.00 & 69.61 \\
		& F1-score (\%) & 61.52 & 45.50 & 81.02 & 14.86 & 57.06 & 69.21 \\
		\cline{3-8}
		& Accuracy (\%) & \multicolumn{6}{c}{69.91} \\
		\midrule
		\multirowcell{4}{RoBERTa} & Precision (\%) & 62.88 & 47.12 & 80.99 & 22.65 & 66.23 & 70.03 \\
		& Recall (\%) & 62.67 & 45.54 & 83.97 & 13.89  & 57.11 & 71.10 \\
		& F1-score (\%) & 62.77 & 46.32 & 82.45 & 17.22 & 61.33 & 70.56 \\
		\cline{3-8}
		& Accuracy (\%) & \multicolumn{6}{c}{71.10} \\
		\midrule
		\multirowcell{4}{CodeBERT} & Precision (\%) & 62.13 & 45.12 & 81.02 & 18.98 & 66.74 & 69.49 \\
		& Recall (\%) & 62.67 & 43.77 & 83.96 & 11.00 & 56.89 & 70.72 \\
		& F1-score (\%) & 62.40 & 44.43 & 82.46 & 13.93 & 61.42 & 70.10 \\
		\cline{3-8}
		& Accuracy (\%) & \multicolumn{6}{c}{70.72} \\
		\bottomrule[1pt]
	\end{tabular}
	\label{rq2result}
\end{table*}

\subsection{RQ2: Effectiveness Comparison}
As shown in Table~\ref{rq2result}, it contrasts the effectiveness of CLeBPI against three existing pre-trained models. We calculate the accuracy (A), precision (P), recall (R), F1-score (F), and weighted average P, R, and F to measure each model.

We first can observe that although these three pre-trained language models are not designed for bug priority inference, they outperform all the baseline approaches in RQ1, which demonstrates that pre-trained language models have powerful context learning abilities. By comparing CLeBPI with BERT, RoBERTa, and CodeBERT, it can obtain the absolute improvements of 6.32\%, 4.97\%, and 5.43\% in terms of weighted average F1-score. In terms of accuracy, CLeBPI gets the absolute improvements by 6.28\%, 5.09\%, and 5.47\% when compared with BERT, RoBERTa, and CodeBERT. These experimental results verify two things: 1) Pre-training CLeBPI on bug report corpus can further improve its representation ability for bug reports, which helps CLeBPI perform more precise bug priority inference; 2) Contrastive learning can help the CLeBPI to learn the deep semantic differences between bug reports, improving the classification performance. Additionally, we can also observe that these three pre-trained language models perform better than all the baseline approaches in RQ1 for addressing the label imbalance problem, which shows that contextual information helps models distinguish bug reports with different priority labels. It also further supports that contrastive learning is helpful for dealing with the label imbalance problem.

\begin{tcolorbox}
	\textbf{Answer to RQ2}: Since we perform a two-stage pre-training for CLeBPI, it can learn the more precise contextual representation of bug reports than our selected three pre-trained language models. Additionally, the success of CLeBPI in addressing the label imbalance problem shows that contrastive learning helps alleviate the label imbalance in the dataset by making CLeBPI learn the deep semantic differences between bug reports with different priority labels.
\end{tcolorbox}

\subsection{RQ3: Retrieval the Effect of bug report's Length} \label{sec53}

\begin{table}[t]
	\centering
	\caption{The accuracy (\%) of CLeBPI on bug reports in different length intervals.}
	\begin{tabular}{l|c|c|c|c|c|c}
		\toprule[1pt]
		\multirow{2}*{-} & \multicolumn{6}{c}{Length of Bug Reports} \\
		\cline{2-7}
		& 0-100 & 100-200 & 200-300 & 300-400 & 400-500 & \textgreater500 \\
		\midrule
		Bug Priority Inference & 76.16 & 76.03 & 75.31 & 78.11 & 79.12 & 77.98 \\
		\bottomrule[1pt]
	\end{tabular}
	\label{rq3result1}
\end{table}

\begin{table}[t]
	\centering
	\caption{The performance of CLeBPI with different settings of the max length of the bug report.}
	\begin{tabular}{c|ccc}
		\toprule[1pt]
		\multirow{2}*{Max length of Bug Reports} & \multicolumn{3}{c}{Bug Priority Inference} \\
		\cline{2-4}
		& Weighted Average F1-score (\%) & Accuracy (\%)& training time/min \\
		\midrule
		64 & 73.77 & 74.25 & 225.66 \\
		128 & 74.98 & 75.43 & 307.12 \\
		256 & 75.53 & 76.19 & 509.56 \\
		512 & 75.54 & 76.22 & 998.43 \\
		\bottomrule[1pt]
	\end{tabular}
	\label{rq3result2}
\end{table}

\noindent\textbf{The effect of bug reports' length} Table~\ref{rq3result1} gives the performance of CLeBPI on the priority inference for bug reports with different length intervals. From the table we can obviously find that CLeBPI achieves more than 75\% accuracy on the priority inference for bug reports with each length interval, which shows that short bug reports have no influence on the performance of CLeBPI. We also find that CLeBPI achieves better performance on the priority inference for bug reports that belong to 400-500 length interval, which is similar to \emph{zhang et al.}'s study \cite{zhang2017bug} that developers take fewer days to fix bugs whose bug reports are between 400 to 500 in length. As the bug reports' length increases (when more than 500), CLeBPI cannot get the performance improvement. We guess the potential reason is that bug reports whose length is more than 500 contain much redundant information, which may be regarded as noise by models.

\noindent{\textbf{The effect of bug reports' max length}} To explore the impact of bug reports' max length on the performance of CLeBPI, we choose four different settings of bug reports' max length when training CLeBPI for bug priority inference, i.e., 64, 128, 256, and 512. As shown in Table~\ref{rq3result2}, it gives the performance of CLeBPI when training with different settings of bug reports' max length. As the bug reports' max length increases, CLeBPI can obtain a higher weighted average F1-score and accuracy. However, we must notice that the training time has a more dramatic increase, thus we need to comprehensively consider the effectiveness and training time. From the table, we observe that when increasing the bug reports' max length from 256 to 512, CLeBPI only obtains a slight improvement. Meanwhile, we can find that the training time raises about 2 times when increasing the max length from 256 to 512. It means that the setting of the max length is not the larger the better. Therefore, we think that setting the max length of bug reports to 256 is a suitable choice that both consider the effectiveness and training overhead.

\begin{tcolorbox}
	\textbf{Answer to RQ3:} To sum up, we confirm two things: 1) the bug reports' length have limited influence on the performance of CLeBPI, which shows its powerful representation learning ability; 2) The setting of bug reports' max length is not the larger the better since it causes an aggressive increase in the training time, thus setting the max length of bug reports to 256 is an appropriate selection.
\end{tcolorbox}

\begin{table}[t]
	\centering
	\caption{The performance comparison of CLeBPI between using contrastive learning and ignoring it.}
	\begin{tabular}{l|cc}
		\toprule[1pt]
		\multirow{2}*{-} & \multicolumn{2}{c}{Bug Priority Inference} \\
		\cline{2-3}
		& Weighted Average F1-score (\%) & Accuracy (\%)  \\
		\midrule
		CLeBPI w/ contrastive learning objective & \textbf{75.53} & \textbf{76.19}  \\
		CLeBPI w/o contrastive learning objective & 72.33 & 72.89 \\
		\bottomrule[1pt]
	\end{tabular}
	\label{ae1}
\end{table}

\begin{table}[t]
	\centering
	\caption{The performance comparison of different data augmentation methods used for contrastive learning.}
	\begin{tabular}{l|cc}
		\toprule[1pt]
		\multirow{2}*{-} & \multicolumn{2}{c}{Bug Priority Inference} \\
		\cline{2-3}
		& Weighted Average F1-score (\%) & Accuracy (\%)  \\
		\midrule
		Masking one word in bug report & 74.84 & 75.47  \\
		Deleting one word in bug report & 74.35 & 74.89 \\
		Exchanging the position of two words in bug report & \textbf{75.53} & \textbf{76.19} \\
		\bottomrule[1pt]
	\end{tabular}
	\label{ae2}
\end{table}

\section{Discussion}\label{section6}
In this section, we first perform some ablation experiments, then specifically discuss the threats to the validity of our experiments.

\subsection{Ablation Experiment}\label{sec61}
\subsubsection{The Impact of Contrastive Learning}\label{sec611}
We explore whether contrastive learning helps CLeBPI improve the effectiveness of bug priority inference. Specifically, we re-run the whole training process of CLeBPI but ignore the second-stage pre-training that uses a contrastive learning objective, then compare its performance with CLeBPI that is conducted the entire pre-training process. As shown in Table~\ref{ae1}, we can find that CLeBPI without contrastive learning performs better than all the baseline approaches in RQ1 and RQ2, which supports that pre-training on bug report corpus is effective. However, we also find that CLeBPI without contrastive learning fails to deal with the label imbalance problem because its F1-score on the priority inference of bug reports with the P4 label is just 13.97\%, which is far lower than the performance of CLeBPI with contrastive learning (i.e., 28.83\%). Additionally, when using contrastive learning to perform a second-stage pre-training for CLeBPI, it gets the absolute improvements of 3.20\% and 3.30\% in terms of weighted average F1-score and accuracy, respectively. All these facts verify the effectiveness of contrastive learning.

We also investigate the impact of different data augmentation methods on the performance of contrastive learning. As shown in Table~\ref{ae2}, we can find that each data augmentation method is helpful for pre-training CLeBPI with a contrastive learning objective, which supports the effectiveness of contrastive learning. We also can observe that the data augmentation method that exchanges the position of two words achieves the highest scores both in weighted average F1-score and accuracy. We guess the potential reason is that exchanging the position of two words in bug reports can produce a more challenging pseudo-positive example, which enforces CLeBPI to learn the deep and useful semantic differences of bug reports, to effectively distinguish them.

\begin{table}[t]
	\centering
	\caption{The performance of CLeBPI on the priority inference for bug reports with different learning rates.}
	\begin{tabular}{l|c|c|c|c|c}
		\toprule[1pt]
		\multirow{2}*{-} & \multicolumn{5}{c}{Learning Rate} \\
		\cline{2-6}
		& 1e-6 & 2.5e-6 & 5e-6 & 7.5e-6 & 1e-5 \\
		\midrule
		Weighted Average F1-score (\%) & 74.16 & 74.77 & \textbf{75.53} & 75.01 & 74.83 \\
		Accuracy (\%) & 74.80 & 75.42 & \textbf{76.19} & 75.77 & 75.36 \\
		\bottomrule[1pt]
	\end{tabular}
	\label{ae3}
\end{table}

\subsubsection{The Impact of Learning Rate}
To investigate the effect of learning rate on CLeBPI, we set five different learning rates and train CLeBPI with them for the bug priority inference. Table~\ref{ae3} gives the performance of CLeBPI when training with different learning rates. Specifically, as the learning rate increases, CLeBPI can get higher scores both in weighted average F1-score and accuracy. When the learning rate is more than 5e-5, the performance of CLeBPI starts decreasing. Considering that the setting of the learning rate only affects the model's performance, setting the learning rate to 5e-6 is a good choice.

\subsection{Threats to Validity}
This paper mainly suffers from some threats to validity. One critical threat to the internal validity is how to effectively set the hyper-parameters in CLeBPI. We mitigate this threat by setting most hyper-parameters according to the prior studies \cite{devlin2018bert, liu2019roberta}, which are verified that these settings are optimal. For other hyper-parameter settings, we also perform sufficient experiments to find the optimal settings, which can be seen in Section~\ref{sec53} and Section~\ref{sec61}. Another threat to the internal validity is that we produce positive instances for bug reports (used for contrastive learning pre-training) by exchanging positions of any two words in bug reports. However, there are other methods to produce positive instances for bug reports and we cannot ensure our selected method is more effective. We mitigate this threat by comparing our selected method with other methods, and we evaluate CLeBPI's performance with different methods, which can be seen in Sec~\ref{sec611}.

One threat to the external validity is that CLeBPI can only serve for bug reports from the Bugzilla platform. There are many different bug tracking systems that also contain lots of bug reports, and they also have a requirement for automated bug priority inference. However, we cannot confirm whether CLeBPI can serve for bug reports from these bug tracking systems because they may have different definitions for the bug priority. A mitigating factor is that CLeBPI can be applied to other bug tracking systems by the transfer learning \cite{weiss2016survey}. Although bug tracking systems have their own definitions for the bug priority, bug reports in these systems also contain Summary and Description elements, which enables CLeBPI to be quickly transformed to these systems by the transfer learning \cite{weiss2016survey}. Specifically, we only need to directly fine-tune CLeBPI on bug reports from the new bug tracking system for the bug priority inference and need not pre-train it again.

\section{Related Work}\label{section7}
In this section, we describe some related studies, including bug priority inference, pre-trained language models, and bug report-related software engineering tasks.

\subsection{Priority Inference}
Bug priority inference is first proposed by \emph{Abdelmoez et al.} \cite{abdelmoez2012bug}, which is an orthogonal task of bug severity prediction \cite{tian2012information, zhang2016towards} and can improve the efficiency of software maintenance by distinguishing the fixing priority of the newly reported bug. Specifically, they utilized na{\"\i}ve Bayes classifier to analyze bug reports for distinguishing the fixing time of their corresponding bugs, by which they can realize the bug priority inference. Afterward, \emph{Alenezi et al.} \cite{alenezi2013bug} compared the performance of na{\"\i}ve Bayesian, random forest, and decision tree on the bug priority inference. Their experimental results show that decision tree and random forest are better than na{\"\i}ve Bayesian. \emph{Tian et al.} proposed DRONE, a linear regression based model that performs bug priority inference by exploiting multiple factors of bug reports. To resolve the label imbalance problem, they extra introduced a thresholding method. Hence, DRONE is the state-of-the-art TML-based approach.

As deep learning \cite{lecun2015deep} becomes popular, NN-based approaches outperform TML-based approaches because neural networks can automatically learn semantic features and need not manual feature engineering. \emph{Choudhary et al.} \cite{choudhary2017neural} utilized multilayer perceptron to perform the bug priority inference, which needs no manual feature engineering and reduces the time-consuming. \emph{Umer et al.} \cite{umer2019cnn} proposed cPur, which is based on a convolutional neural network and can learn the local semantic information of bug reports, outperforming DRONE on the bug priority inference. \emph{Bani-Salameh et al.} \cite{bani2021deep} proposed a novel approach for the bug priority inference, which is built by stacking five layers of RNN-LSTM neural networks. \emph{Fang et al.} \cite{fang2021effective} proposed PPWGCN, a graph convolutional network based approach than introduces a weighted loss function to deal with the label imbalance problem, thus it becomes the state-of-the-art NN-based approach for the bug priority inference.

Different from the above-mentioned approaches, we first pre-train CLeBPI with a masked language model objective, to learn the contextual representation of bug reports. Then, we further pre-train CLeBPI with a contrastive learning objective, by which CLeBPI can learn the deep semantic differences between bug reports. Finally, we train CLeBPI for the bug priority inference, which achieves state-of-the-art results on the bug priority inference and can effectively alleviate the label imbalance problem.

\subsection{Pre-trained Language Model in Software Engineering}
Pre-trained language models have been widely utilized to perform other software engineering tasks. For example, \emph{Feng et al.} \cite{feng2020codebert} proposed CodeBERT, which can learn contextual representation of code and natural language, achieving state-of-the-art results both in code search \cite{fang2021self} and code comment generation tasks \cite{hu2018deep}. \emph{Jiang et al.} proposed CURE, which is a neural machine translation based automated program repair approach. Before performing the automated program repair, it is pre-trained on a large-scale software codebase to fully learn the contextual information of the source code. \emph{Liu et al.} \cite{liu2020multi} proposed CugLM, a Transformer-based pre-trained language model pre-trained with a hybrid objective, containing both code understanding and code generation. Their approach thus brings new results to code completion tasks on two public datasets.

Different from the aforementioned approaches, CLeBPI is pre-trained to learn the contextual representation of bug reports and is used for the bug priority inference. Moreover, to deal with the label imbalance problem, we introduce contrastive learning and perform a two-stage pre-training for CLeBPI, by which we effectively alleviate the label imbalance problem while improving the performance of CLeBPI.

\subsection{bug report-Related Software Engineering Tasks}
Except for the bug priority inference, there are other bug report-related automated software engineering tasks \cite{he2020duplicate, mani2012ausum, tan2020bug, xia2016improving, zhou2012should}. For example, duplicate bug report detection \cite{jalbert2008automated} can help developers find bugs that have been reported, improving their efficiency. Bug report summarization can automatically generate the titles for bug reports, which helps fixers quickly understand the bug and perform effective fixing. As for the bug severity prediction \cite{zhang2016towards}, it can tell the severity of the newly submitted bug, and bug traiger can perform the reasonable assignment. Automated bug triaging \cite{bhattacharya2010fine} can assign a newly submitted bug to the appropriate developer to fix it, which can increase the fixing rate and speed of bugs. Bug localization \cite{liu2005sober} can help developers quickly locate where the bug appears, improving the fixing speed of bugs. To sum up, all bug report-related tasks can help improve the efficiency of software maintenance.

\section{Conclusion}\label{section8}
In this paper, we focus on improving the performance of bug priority inference based on bug reports, including effectiveness and addressing the label imbalance problem in the dataset. To achieve our goal, we propose CLeBPI, a Transformer-based pre-trained language model. To improve the effectiveness of CLeBPI, we pre-train it with a masked language model objective to learn the precise contextual representation of bug reports. To cope with the label imbalance problem, we perform a second-stage pre-training for CLeBPI with a contrastive learning objective, to learn the deep semantic differences between bug reports. Finally, we train CLeBPI for the bug priority inference and experimental results show that it outperforms all baseline approaches and effectively alleviates the label imbalance problem.

In the future, we plan to further explore the label imbalance problem by diving into bug reports with rare priority labels, to find a more effective approach to resolve it. Additionally, we will also plan to explore the generalizability of CLeBPI by applying it to other bug report-related software engineering tasks.


\section*{Acknowledgment}
This work was supported in part by the National Natural Science Foundation of China (Grant No. 61961036, 62162054), in part by Natural Science Foundation of Guangxi (Grant No. 2020JJA170007), and in part by Special Project of Guangxi Science and Technology Base and Talent (Guike AD20297148).


\bibliographystyle{cas-model2-names}

\bibliography{ref}

\end{document}